\documentstyle[prl,aps,epsbox]{revtex}
\begin{document}
\draft
\twocolumn[\hsize\textwidth\columnwidth\hsize\csname@twocolumnfalse%
\endcsname 
\title{ Realizing Discontinuous Wave Functions 
with Renormalized Short-Range Potentials }
\author{
Taksu Cheon
}
\address{
The Laboratory of Physics, Kochi University of Technology,
Tosa Yamada, Kochi 782-8502, Japan \\
email: cheon@mech.kochi-tech.ac.jp,
http://www.kochi-tech.ac.jp/\~{}cheon/
}
\author{
T. Shigehara
}
\address{
Department of Information and Computer Sciences,  
Saitama University,
Urawa, Saitama 338-8570, Japan \\
email: sigehara@ics.saitama-u.ac.jp, 
http://www.me.ics.saitama-u.ac.jp/\~{}sigehara/
}

\vspace*{2mm}
\date{February 23, 1998}
\maketitle
\begin{abstract}
We show that the most general three-parameter family of point 
interactions on the line can be expressed as 
the self-adjoint local operators in terms of three Dirac's 
delta functions with the renormalized strengths in the 
disappearing distances.
Experimental realization of the Neumann boundary is discussed.

\vspace*{2mm}

\end{abstract}
%
]

%
%

The discontinuity of the gradient of wave functions has been known 
since the very early days of quantum mechanics.
Even today, however, 
the admissibility of the discontinuity in the wave function
itself is not well recognized outside
the circle of mathematical physicists.
This is mainly due to the esoteric languages employed in formulating
the subject, which tends to give an impression that the phenomena are
far removed from the experimentally realizable settings.

Let us focus on an example of the simplest, but non-trivial 
wave function discontinuity that can be found in the 
free particle quantum mechanics on one-dimensional line 
with a pointlike interaction, 
or a point defect \cite{GH80,GK85,SE86,SE86a,GH87,AG88}.  
Assuming the time-reversal
symmetry, the effect of the interaction is usually expressed in terms
of the three parameter connection condition between the wave
functions and their derivatives at the left and the right of the
location of the defect.  Depending on the parameter value, the
condition allows the discontinuities both of wave function and its
space-derivative.  In addition,  there can be left-right asymmetry in the
connection condition at the defect.  In
the abstract mathematical treatment, however, 
no specific prescription is given for the
realization of this connection condition as Hamiltonian dynamics.
Despite several attempts for the physical 
interpretation \cite{AE94,KI97}, the intuitive picture is still lacking.
That is evident, for example, in the controversy over its appellation of
``$\delta'$ interaction'' (see the discussions
in \cite{EX96,CN97}).  Also, no experiment can be conceived to check and
utilize their effect.  

Two recent works \cite{CA93,CH93}
on the potential models whose zero-range limit give the full
connection condition go some way to address this problem.  But the
potentials obtained there are non-local and non-Hermitian
except in the zero-range limit, thus still leave us puzzled over
their relevance to the real-world quantum mechanics.  It would be very
useful if we can express the point defect in terms of a zero-range
limit of well behaving functions.

The objective of this paper is to devise just such potential
functions.  
It is done in terms of zero-distance limit of three or more Dirac's
delta functions.  This effectively gives the practical
prescription to realize the wave function discontinuity as well as
its intuitive understanding, since, the delta function in one
dimension, in tern, is realizable as a regular limit of small size
potential with the volume integral kept constant. 

%
%

We start by defining a function using three Dirac's
delta functions placed on the line separated by small distances
$a$;  
\begin{eqnarray} 
\label{xidf} 
\xi (x;v,u,a)=v\delta (x+a)+u\delta(x)+v\delta (x-a). 
\end{eqnarray}
The strengths of the delta functions are allowed to be varied as
functions of the distance $a$; 
\begin{eqnarray}
\label{funct}
v = v(a) , \ \ \ u = u(a) \ \ \ {\rm as} \ \ \ a \to 0.
\end{eqnarray}
We look at the quantum-mechanical wave function $\psi (x)$
under the influence of the potential $\xi(x;v,u,a)$ 
\begin{eqnarray}
\label{schro}
-{1 \over 2}{{d^2} \over {dx^2}}\psi (x)+\xi (x;v,u,a)\psi (x)
= E\psi (x).
\end{eqnarray}
The solution of the Schr{\" o}dinger equation in the intervals
$x\in [0_+,a_-]$,  where $x_-$ and $x_+$
respectively signify the location infinitesimally smaller and 
larger than $x$, can be written in the form
\begin{eqnarray}
\label{2cond1}
\psi(x) &=& \psi (0) \cos {k x} + \psi' (0_+) k^{-1} \sin k x , 
\\ \nonumber
\psi'(x) &=& \psi' (0_+) \cos {k x} - \psi (0) k \sin k x , 
\end{eqnarray}
where $k$ is the wave number defined by $k \equiv \sqrt{2 E}$.  
Assuming that we are interested in the 
low energy spectra of Eq. (\ref{schro}), 
we expand Eq. (\ref{2cond1}) in terms of $k$ to take the leading order,
and obtain
\begin{eqnarray}
\label{1conda}
\psi (a) &=& \psi (0)+a\psi' (0_+) ,
\\ \nonumber
\psi' (a_-) &=& \psi' (0_+) .
\end{eqnarray}
With analogous treatment for $x\in [(-a)_+,0_-]$, we have
\begin{eqnarray}
\label{1condb}
\psi (-a) &=& \psi (0)-a\psi' (0_-) ,
\\ \nonumber
\psi' ((-a)_+) &=& \psi' (0_-) .
\end{eqnarray}
Note that the wave function $\psi(x)$ itself is continuous everywhere
at this stage.
Combining Eqs. (\ref{1conda}) and (\ref{1condb}) 
with the connection conditions at the location 
of the three Dirac's deltas,
\begin{eqnarray}
\label{5cond}
\psi' ((-a)_-)-\psi' ((-a)_+) &=& -2v\psi (-a) , 
\\ \nonumber
\psi' (0_+)-\psi' (0_-) &=& 2u\psi (0) , 
\\ \nonumber
\psi' (a_+)-\psi' (a_-) &=& 2v\psi (a) ,
\end{eqnarray}
we obtain 
\begin{eqnarray}
\label{diffps}
\psi' (a_+)-\psi' ((-a)_-)
&=& B(a) \left\{{\psi (a)+\psi (-a)} \right\} ,
\\ \nonumber
\psi (a)-\psi (-a)
&=& D(a) \left\{ {\psi' (a_+)+\psi' ((-a)_-)}\right\} ,
\end{eqnarray}
where the functions $B(a)$ and $D(a)$ are given by
\begin{eqnarray}
\label{diffbd}
B(a) &=& 2v(a)+{{u(a)} \over {1+au(a)}} , 
\\ \nonumber
D(a) &=& {a \over {2av(a)+1}} .
\end{eqnarray}
We now take the limit $a \to 0$ and look at the relations between
the quantities
\begin{eqnarray}
\label{a0lim}
\psi_\pm 
&\equiv& \mathop {\lim}\limits_{a\to 0}\psi(\pm a) ,
  \\ \nonumber
\psi'_\pm
&\equiv& \mathop {\lim}\limits_{a\to 0}\psi'((\pm a)_\pm) ,
\end{eqnarray}
which express the possible discontinuity of
the wave function and its derivative at the location of the
interaction which is now a single point, $x = 0$. 

We can choose the function $v(a)$ and $u(a)$ so that 
either of $B(a)$ or $D(a)$ becomes zero at $a \to 0$
limit while keeping the other finite. One obvious choice
is to keep $v(a)$ and $u(a)$ constant, that is
\begin{eqnarray}
\label{choice1}
v(a) = v_0, \ \ \ u(a) = u_0 .
\end{eqnarray}
This results in a potential function
\begin{eqnarray}
\label{dfdelt}
\delta (x;2v_0+u_0)
\equiv \mathop {\lim}\limits_{a\to 0}\xi (x;v_0,u_0,a)
\end{eqnarray}
which yields the conditions
\begin{eqnarray}
\label{deltc}
\psi'_+-\psi'_- &=& 2 (2v_0+u_0)\psi _- ,
  \\ \nonumber
\psi _+-\psi _- &=& 0 .      
\end{eqnarray}
This means 
\begin{eqnarray}
\label{deltm}
\delta(x;v) = v \delta(x)      
\end{eqnarray}
which is a rather trivial result.

Now comes the second choice, which is obtained by setting 
$v(a) \to -1/(2a)$ to make the $D(a)$ non-zero and
choosing $u(a)$ so that $B(a)$ becomes zero;  namely
\begin{eqnarray}
\label{choice2}
v(a) &=& {1 \over {2c}}-{1\over {2a}} \\ \nonumber
u(a) &=& -{1 \over a}+{c \over {a^2}} .
\end{eqnarray}
We then have a potential
\begin{eqnarray}
\label{dfepsi}
\varepsilon (x;c) 
\equiv \mathop {\lim}\limits_{a\to 0}
\xi (x;{1 \over {2c}}-{1\over {2a}},
     -{1 \over a}+{c \over {a^2}},a)
\end{eqnarray}
which results in
\begin{eqnarray}
\label{epsic}
\psi'_+ - \psi'_- &=& 0 ,
  \\ \nonumber
\psi _+ - \psi _- &=& 2c\psi'_- ,      
\end{eqnarray}
which is what we wanted; the discontinuity of the wave function.

Several remarks on the discontinuity-inducing
potential $\varepsilon(x;c)$ are
in order.  In our construction, the function is even with respect
to the transposition $x \leftrightarrow -x$.  This fact does not
seem to favour its interpretation in terms of ``$\delta'(x)$'' 
as has been done in the literature. The existence
of the third delta in the middle, which is essential to get the
convergence, also seems to preclude the interpretation of this
function in terms of difference between two deltas.  An important
special case of $\varepsilon(x;c)$ is obtained in the limit 
$c \to \infty$, which results in $\psi'(0) = 0$.  
This is non other than the
{\it Neumann boundary condition} that separates the system into 
the two regions $x > 0$ and $x < 0$.  This is contrasted to 
the Dirichlet boundary condition, which is of course obtained 
as the $v=\infty$ limit of $\delta(x;v)$.

%
%
%

The self-adjoint extension theory applied to the quantum mechanics
on the line with a point defect gives 
the most general connection condition for the wave function at the
site of defect in the form  \cite{AG88}
\begin{eqnarray}
\label{dfmat}
  \left( {\matrix{{\psi'_+}\cr{\psi _+}\cr}} \right)
= \left( {\matrix{{-\alpha }&{-\beta }\cr
                 {-\delta }&{-\gamma }\cr}} \right)
  \left( {\matrix{{\psi'_-}\cr{\psi _-}\cr}} \right)
\end{eqnarray}
with a constraint
\begin{eqnarray}
\label{cnstrn}
\alpha \gamma - \beta \delta = 1 .
\end{eqnarray}
The special choice 
$\alpha = \gamma =-1$ gives the conditions
eq. (\ref{deltc}) (for $\delta=0$) and eq. (\ref{epsic}) (for
$\beta=0$).  We therefore have the correspondence
\begin{eqnarray}
\label{corrsp}
\left( {\matrix{{1}&{2v}\cr{0}&{1}\cr}} 
\right)
&\longleftrightarrow& \delta(x;v) ,
\\ \nonumber
\left( {\matrix{{1}&{0}\cr{2c}&{1}\cr}} 
\right)
&\longleftrightarrow& \varepsilon(x;c) .
\end{eqnarray}

In order to obtain the potential function which gives the full
connection condition, 
one can work out in a similar fashion as before,
with different choice of $v(a)$ and $u(a)$, introducing different
strengths to the deltas of left and right.
But we take different route here, which is
easier and, in a sense, gives us better insight into the meaning of
the parameter values $\alpha$ $\beta$ $\gamma$ and $\delta$.

The method is based on the matrix identity
\begin{eqnarray}
\label{matmlt}
\left( {\matrix{{-\alpha }&{-\beta }\cr
       {-\delta }&{-\gamma }\cr}} 
\right)
=\left( {\matrix{1&{\frac{\alpha +1}{\delta} }\cr 0&1\cr}} 
 \right)
 \left( {\matrix{1&0\cr{-\delta }&1\cr}} 
 \right)
 \left( {\matrix{1&{ \frac{\gamma +1}{\delta} }\cr 0&1\cr}} 
 \right) .
\end{eqnarray}
Successive application of the matrix to the vector 
$(\psi', \psi)$ can be implemented as the point interactions
placed next to each other in disappearing distance.
With the correspondence eq. (\ref{corrsp}) in mind, it is 
easy to convince oneself that the potential function  
\begin{eqnarray}
\label{dfkai}
& &\chi (x;\alpha ,\beta ,\gamma ,\delta )
\\ \nonumber
&\equiv& \mathop {\lim}\limits_{b\to 0}
  \left\{ {\delta(x+b;{{\gamma +1} \over {2\delta }})
          +\varepsilon (x;{-\delta \over 2})
          +\delta (x-b;{{\alpha +1} \over {2\delta }})} 
  \right\}
\end{eqnarray}
produces the desired connection condition, eq. (\ref{dfmat}) at
$x = 0$.  Since no singularity in the coupling is present in the
limit $b \to 0$, one can make $b$ arbitrarily small as long
as it doesn't contradict with the definitions eqs.(\ref{dfdelt}) and 
(\ref{dfepsi}).  This means that one can "merge" the two deltas at 
$x = -b$ and $x = b$ into the two deltas at $x = -a$ and $x = a$ 
which constitute peripheral flank of 
$\varepsilon(x;c)$.  One arrives at
\begin{eqnarray}
\label{dfkai3d}
& &\chi (x;\alpha ,\beta ,\gamma ,\delta )
\\ \nonumber
&=& \mathop {\lim}\limits_{a\to 0}
  \left\{  
    {
    \delta(x+a;{{\gamma-1} \over {2\delta }}-{1 \over {2a}})
   +\delta(x;-{1 \over {a}}-{\delta \over {2a^2}})
    }
  \right. \\ \nonumber
& & \ \ \ \ 
  \left.
   {
   +\delta(x-a;{{\alpha-1} \over {2\delta }}-{1 \over {2a}})
    } 
  \right\}.
\end{eqnarray}
It is not difficult to confirm this expression with direct
calculation similar to eqs. (\ref{5cond})-(\ref{diffps}). 

The expressions eqs. (\ref{matmlt}) - (\ref{dfkai3d}) are not valid
for $\delta = 0$, in which case we resort to the expression

\begin{eqnarray}
\label{matmlt2}
\left( {\matrix{{-\alpha }&{-\beta }\cr
       {-\delta }&{-\gamma }\cr}} 
\right)
=\left( {\matrix{1&0\cr {\frac{\gamma +1}{\beta} }&1\cr}} 
 \right)
 \left( {\matrix{1&{-\beta }\cr 0&1\cr}}
 \right)
 \left( {\matrix{1&0\cr {\frac{\alpha +1}{\beta} }&1\cr}} 
 \right) ,
\end{eqnarray}
which, in terms of the potential function, means
\begin{eqnarray}
\label{dfkai2}
& &\chi (x;\alpha ,\beta ,\gamma ,\delta )
\\ \nonumber
&=& \mathop {\lim}\limits_{b\to 0}
  \left\{ {\varepsilon (x+b;{{\alpha +1}\over {2\beta }})
          +\delta (x;{-\beta  \over 2})
          +\varepsilon (x-b;{{\gamma +1}\over {2\beta }})} 
  \right\} .
\end{eqnarray}
Combining this and eqs.(\ref{dfdelt}) and
(\ref{dfepsi}), and again ``merging'' the neighbouring deltas where
appropriate, we obtain an expression which involve five deltas;
\begin{eqnarray}
\label{dfkai5d}
& &\chi (x;\alpha ,\beta ,\gamma ,\delta )
\\ \nonumber
&=& \mathop {\lim}\limits_{a\to 0}
  \left\{  
   {
    \delta(x+2a;{{\beta}\over {\alpha+1}}-{1 \over {2a}})
   +\delta(x+a;-{1 \over {a}}+{{\alpha+1} \over {2\beta a^2}}) 
   }
  \right.
\\ \nonumber
& &\ \ \ \ 
   +\delta(x;{{\beta}\over{\alpha+1}}+{{\beta}\over{\gamma+1}}
             -{{\beta}\over{2}}
             -{1 \over {a}} )
\\ \nonumber
& &\ \ \ \ 
  \left.
   {
   +\delta(x-a;-{1 \over {a}}+{{\gamma+1} \over {2\beta a^2}})
   +\delta(x-2a;{{\beta}\over {\gamma+1}}-{1 \over {2a}})
   }
  \right\}.
\end{eqnarray}

The expressions eqs. (\ref{matmlt2}) - (\ref{dfkai5d}) are still not
applicable to the special case of $\beta = \delta = 0$.
For this case, we use another expression
\begin{eqnarray}
\label{matmlt3}
\left( {\matrix{{-\alpha }&0\cr
       {0 }&{-\gamma }\cr}} 
\right)
&=&\left( {\matrix{0&\rho\cr {-\frac{1}{\rho}}&0\cr}} 
 \right)
 \left( {\matrix{0&{\mp\frac{1}{\rho}}\cr \pm\rho&0\cr}} 
 \right)
\\ \nonumber
&=&\left( {\matrix{1&\rho\cr 0&1\cr}} 
 \right)
 \left( {\matrix{1&0\cr {-\frac{1}{\rho}}&1\cr}} 
 \right)
 \left( {\matrix{1&\rho\cr 0&1\cr}} 
 \right)
\\ \nonumber
&\times&
 \left( {\matrix{1&{\mp\frac{1}{\rho}}\cr 0&1\cr}} 
 \right)
 \left( {\matrix{1&0\cr {\pm\rho}&1\cr}} 
 \right)
 \left( {\matrix{1&{\mp\frac{1}{\rho}}\cr 0&1\cr}} 
 \right) ,
\end{eqnarray}
where $\rho \equiv \sqrt{|\alpha|}= 1/\sqrt{|\gamma|}$ and the composite
sign corresponds to the case of $\alpha = \mp |\alpha|$.
To obtain the second equality, eq. (\ref{matmlt}) is applied
to the both matrices in the RHS of the first equality. 
One can implement this expression in potential form as
\begin{eqnarray}
\label{dfkai5z}
& &\chi (x;\alpha ,0 ,\gamma ,0 )
\\ \nonumber
&=& \mathop {\lim}\limits_{a\to 0}
  \left\{  
   {
    \delta(x+2a;\pm{1 \over {2 \rho}}-{1 \over {2a}})
   +\delta(x+a;-{1 \over {a}}\pm{{\rho} \over {2 a^2}})
   }
  \right.
\\ \nonumber
& &\ \ \ \ 
   +\delta(x; -{\rho \over 2} \pm{1 \over {2\rho}}
             -{1 \over {a}} )
\\ \nonumber
& &\ \ \ \ 
  \left.
   {
   +\delta(x-a;-{1 \over {a}}-{{1} \over {2\rho a^2}}) 
   +\delta(x-2a;-{\rho\over {2}}-{1 \over {2a}})
   }
  \right\}.
\end{eqnarray}
Thus, the most general connection condition around the point 
defect in one-dimensional quantum mechanics can be realized as
singular, but renormalized zero-distance limit of three (or
five) Dirac's delta functions. 
We note that the above expressions are by no means unique; 
one can construct expressions with three deltas in place of
eqs. (\ref{dfkai5d}) and (\ref{dfkai5z}) 
with direct method mentioned before. 
But more important than the numerical economy is the fact that the
expressions eqs. (\ref{dfkai}), (\ref{dfkai5d}) and (\ref{dfkai5z})
make the meaning of the asymmetry for $\alpha \neq \gamma$ case 
very clear; one obtains different results by placing $\delta(x;v)$
to the left or to the right of $\varepsilon(x;c)$ even in the 
zero-distance limit.
%
%

We offer a numerical example to illustrate the workings of
the realization of the wave function discontinuity.  
We use finite potential of range $s$ in
place of delta as the building block;
\begin{eqnarray}
\label{dfDelt}
\Delta_s (x) &=& \left\{
\begin{array}{cl} 
\displaystyle 
\frac{2}{s}\cos^2(\frac{\pi}{s}x), &  
 \ \ \ \displaystyle x < \left| s/2 \right|, \\ 
0, & 
 \ \ \ \displaystyle x > \left| s/2 \right|.
\end{array} \right.
\end{eqnarray}
Then, the potential $E_{a,s}(x;c)$ defined by 
\begin{eqnarray}
\label{dfEps}
E_{a,s}(x;c)
&=&  (\frac{1}{2c}-\frac{1}{2a})\Delta_s(x+a)
\\ \nonumber
& &
 + (-\frac{1}{a}+\frac{c}{a^2})\Delta_s(x)
 + (\frac{1}{2c}-\frac{1}{2a})\Delta_s(x-a) 
\end{eqnarray}
becomes a good approximation to, and ultimately converge
toward the $\varepsilon(x;c)$ as one takes the limit
$ 0 < s \ll a \to 0$.

In Fig. 1, the first four eigenstates of the Schr{\" o}dinger 
equation on a line $x \in [-L/2,L/2]$ 
with the potential eq. (\ref{dfEps}) are shown along 
with the potential itself.
Dirichlet conditions are imposed at the edge, namely,
$\psi(L/2) = \psi(-L/2) = 0$.
The value of the coupling is set to be $c = 5$.
The parameters $a$ and $s$, which are supposed to be
smaller than the scale of the problem $L$ (which we
arbitrarily set $L=10$) are chosen to be $a = 0.0333 L$ and  
$s = 0.0012 L$. At this level of ``small but finite'' $s$ and $a$,
one can  already observe the discontinuity of the wave function 
developing around $x=0$ for the second and the fourth states,
while the first and the third states show ``continuity''
because one has $\psi'(0) = 0$ for these even-parity states.  
The near degeneracy found between 
the first and the second states, and also
between the third and the fourth states, can be thought of as the
sign of the closeness to the Neumann limit $c=\infty$ where the
system is divided into the two isolate identical subsystems at
$x=0$.
\begin{figure}
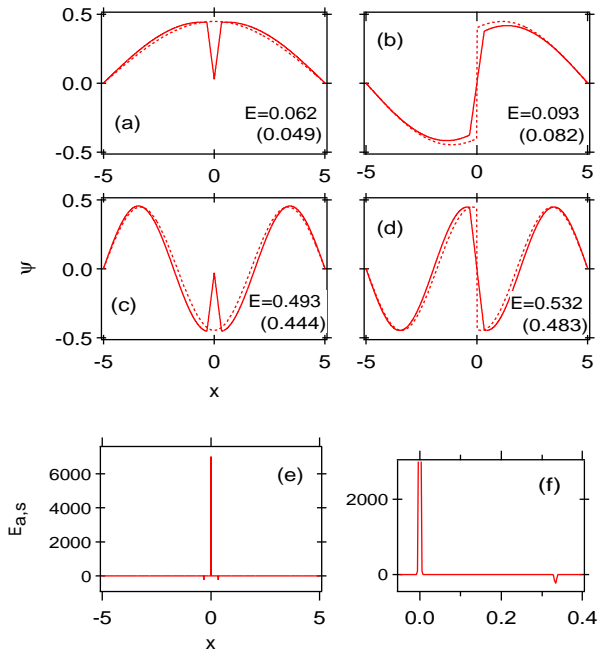

\center\psbox[hscale=0.5,vscale=0.5]{epfig1.epsf}
\caption{
(a) -(d) show the first four eigenstates of the particle of unit
mass on the line $x \in [-L/2,L/2]$ with the
potential defined by eqs. (26) and (27), which is depicted
in (e). The numbers are the energy eigenvalues.
Parameters are $c=5$, $L=10$, $a=0.0333L$, $s=0.0012L$.
The dashed lines show the wave functions (and the numbers in the
brackets the eigenvalues) of the predicted $a \to 0$ limit
calculated from the connection condition eq. (14).  (e) and (f)
depict the potential function in different scales. 
} 
\label{fig1}
\end{figure}

This example clearly shows
that our procedure to realize the wave function discontinuity 
is not a mere mathematical abstraction, but something
{\it actually realizable in experiments}.
Recent progress in the quantum device of nanometer scale  
offers a possible opportunity. 
We also note that the experiments could be carried out with
electro-magnetic and other macroscopic waves, since the results
obtained here is applicable to larger classes of
linear wave equations other than the Schr{\"o}dinger equation.
%

Finally, we place our findings in broader context.
Throughout this paper, we have kept our arguments in the language
of elementary quantum mechanics.  It is not difficult,
however, to reformulate the problem in terms of Green's
function \cite{GR95,PA96}.
In any case, we believe that our procedures clearly show that
there are non-trivial, but {\it experimentally accessible} 
zero-range forces in one 
dimensional quantum mechanics other than the familiar 
Dirac's delta function.

In hindsight, it is natural to expect the existence of certain
singularity, or the divergence  in order to obtain the wave
function discontinuity, since the ordinary delta function is known
to be ``too week'' for that purpose.
For the theory to be still well-defined with divergent quantity,
the renormalization has to be introduced.  
This is in a sense analogous to the situation in two and three
dimensions where one encounters divergence and renormalization in
order to define proper point interactions (see, for example,
\cite{SC97} and references therein).  We have to add in rush that
there is an essential dissimilarity; In two and three dimensions,
one can define only one class of zero-range potential 
which corresponds to $\delta$ function, and there is no
analogue to the function $\varepsilon(x;c)$.

It would be useful to compare the current realization of 
discontinuity-inducing potential to the earlier approach found in
ref \cite{SE86a} which utilizes the
{\it separable potential} \cite{YY54} of infinite rank, since the 
usual $\delta$ function
can be thought of as a separable potential of rank one.

The function $\chi(x;\alpha,\beta,\gamma,\delta)$ belongs to a
solvable class of quantum potential while retaining the richness of
three parameter dependence.
Already, several unusual physical properties are predicted for the
systems with discontinuity-inducing potentials \cite{AE94,KI97}.
Important fact to recall is that {\it all} imaginable potentials
in one dimension are guaranteed to converge toward 
$\chi(x;\alpha,\beta,\gamma,\delta)$ in zero-range limit.
As such, it should help us
unveil non-trivial aspects of simple quantum systems.

In theoretical treatment of quantum system, the Neumann boundary
condition appears as natural as, and at times, more convenient than
the Dirichlet boundary. We hope the current work may stimulate
the experimental design to materialize and utilize the Neumann
boundary condition.

\vspace*{5mm}

We thank Prof. Izumi Tsutsui for helpful discussions.

\end{document}